\documentclass{elsarticle}
\usepackage{amssymb,amsmath,amsfonts,mathrsfs}
\usepackage[utf8]{inputenc}

\usepackage{lineno,hyperref}
\usepackage{graphicx}
\modulolinenumbers[5] 

\biboptions{authoryear}

\newtheorem{remark}{Remark}
\newtheorem{proposition}{Proposition}

\newcommand{\proofofref}{}
\newproof{zproofof}{Proof of \proofofref}

\numberwithin{equation}{section}
\numberwithin{lemma}{section}
\numberwithin{theorem}{section}
\numberwithin{corollary}{section}
\numberwithin{remark}{section}

\begin{document}

\begin{frontmatter}
\title{ A computational approach to  the Kiefer-Weiss problem for sampling from a Bernoulli population}

\author[UAMaddress]{Andrey Novikov\corref
{mycorrespondingauthor}}\ead{an@xanum.uam.mx}
\author[RFaddress]{Andrei Novikov}
\ead{A.HOBuKOB@gmail.com}
\author[UAMaddress]{Fahil Farkhshatov}
\ead{failfarkhshatov@gmail.com}

\cortext[mycorrespondingauthor]{Corresponding author}
\address[UAMaddress]{UAM-Iztapalapa, San Rafael Atlixco 186, col. 
Vicentina, C.P. 09340, Mexico City, Mexico} 
\address[RFaddress]{Sobolev Institute of Mathematics, Siberian Branch of Russian Academy of Sciences, 4 Acad. Koptyug avenue, 630090, Novosibirsk, Russia}

\begin{abstract}

We present a computational approach to solution of the Kiefer-Weiss problem. 

Algorithms for construction of the optimal sampling plans and evaluation  of their performance are proposed. In the particular case of Bernoulli observations, the proposed algorithms are implemented in the form of R program code.

Using  the developed computer program,
we numerically compare the optimal tests with the respective sequential probability ratio test (SPRT) 
and the fixed sample size test, for a wide  range  of  hypothesized  values  and  type I  and type II errors.

The results are compared with those of D.~Freeman and L.~Weiss (Journal of the American Statistical Association,  59(1964)).

The R source code for the algorithms of construction of optimal sampling plans and evaluation of their characteristics is available at \\
{\tt https://github.com/tosinabase/Kiefer-Weiss}.
\end{abstract}

\begin{keyword}
Sequential analysis\sep
hypothesis testing\sep
optimal stopping\sep
optimal sequential tests\sep
Kiefer-Weiss problem\sep
Bernoulli trials
\MSC[2010] 62L10, 62L15, 62F03, 60G40, 62M02
\end{keyword}
\end{frontmatter}









\section{\normalsize  Introduction}\label{sIntro}
In a sequential statistical  experiment, a sequence of random variables $X_1,X_2$, $\dots, X_n, \dots$ is potentially  available to the statistician on the one-by-one basis. The observed data bear the  information  about  the underlying distribution $P_\theta$, being  $\theta$  an unknown parameter whose true value  is of interest to the statistician. In this paper, we are concerned with testing  a simple hypothesis H$_0:$ $\theta={\theta_0}$ against a simple alternative H$_1:$ $\theta={\theta_1}$, which is a classical problem of sequential analysis \citep[see, e.~g.,][]{WaldWolfowitz}.

In its simplest form,
a sequential hypothesis test is a pair $\langle \tau, \delta\rangle$ consisting of a the stopping time  $\tau$  and a (terminal) decision rule $\delta$. Formally, it is required that $\{\tau= n\}\in \sigma(X_1,\dots, X_n)$ and $\{\tau=n,\delta=i\}\in\sigma(X_1,\dots, X_n)$, for any natural $n$ and $i=0,1$.
The performance characteristics of  a sequential test are the type I and type II error probabilities,
$\alpha(\tau,\delta)=P_{\theta_0}(\delta=1)$ and $\beta(\tau,\delta)=P_{\theta_1}(\delta=0)$,
and the average sample number, $E_\theta \tau$.

The Kiefer-Weiss problem is to find a test $\langle\tau,\delta\rangle$ with a minimum value  of 
$\sup_\theta E_\theta \tau$, 
among all the tests satisfying the constraints on the type I and type II error probabilities:  
\begin{equation}\label{2}
\alpha(\tau,\delta)\leq\alpha\quad\mbox{and}\quad\beta(\tau,\delta)\leq\beta.
\end{equation}
\cite{KieferWeiss} noted that
in some cases the solution of this problem can be obtained through the solution of a much simpler one (known, at the time being, as the modified Kiefer-Weiss problem).
To be more specific, suppose that there exists a test $\langle\tau,\delta\rangle $ that minimizes $E_{\theta_*}\tau$, for some fixed $ \theta_*$, among all the tests satisfying \eqref{2},  and  that this $\theta_* $ is the ``least favorable" for the stopping time $\tau$ in the sense that 
\begin{equation}\label{3}
\sup_{\theta}E_{\theta} \tau=E_{\theta_*}\tau.
\end{equation}
Then it is easy to see that $\langle\tau,\delta\rangle $ minimizes $\sup_\theta E_\theta \tau$, among all the tests satisfying \eqref{2}, that is, it solves the original Kiefer-Weiss problem.

In this way,  \cite{Weiss} used Bayesian approach for the solution of the Kiefer-Weiss problem, with $\alpha=\beta$, for two special cases:\\
1) the observations are independent and identically distributed (i.i.d.) with a normal $\mathscr N(\theta,\sigma^2)$ distribution with an unknown mean  $\theta$ and a known variance $\sigma^2$, and
2) the observations are i.i.d. with a Bernoulli distribution  with an unknown success probability $\theta$,
under the additional assumption that $\theta_0=1-\theta_1$.
In neither case the Bayesian test was  easy  to find. Even the simplest Bernoulli case ``requires heavy computing'' (as stated by  \cite{Weiss}, p. 565).

 \cite{FreemanWeiss} applied the same technique to a more general case of the Bernoulli model with non-symmetric hypotheses; they proposed a scheme for finding an approximate numerical solution to the Kiefer-Weiss problem by making $E_{\theta_*}\tau$ close enough to $\sup_\theta E_\theta\tau$ (cf. \eqref{3}).

\cite{Lorden}  characterized  the structure of the optimal tests in the modified Kiefer-Weiss problem for the particular case of one-parametric Koopman-Darmois families of distributions and showed that, for $\theta_*\in (\theta_0,\theta_1)$, the stopping times of these tests are almost surely bounded, and gave this bound.

Many other results concerning  approximate or asymptotic solutions of  the Kiefer--Weiss
problem, and especially the modified Kiefer--Weiss problem, can be found in the literature.
The modern viewpoint on the Kiefer-Weiss problem, and related,  can be found in the monograph  \cite{Tartakovsky2014}.
With respect to the exact solutions to the Kiefer-Weiss problem, which is the point  of our interest in this paper, the authors state that  ``finding the exact solution involves quite heavy computation'' \citep[][p. 228]{Tartakovsky2014}, which is  generally concurring with the opinions expressed earlier by  \cite{Weiss}, \cite{ FreemanWeiss}, and \cite{Lorden}, among others, with respect to the particular cases they studied.

In this paper, we propose a computational approach to the solution of the Kiefer-Weiss problem, which seems to be  general enough to provide an exact solution to virtually any particular case of the Kiefer-Weiss problem, but generally does not require extreme computational power and is within the reach of modern computer capabilities.

We call this approach ``computational'', because it essentially relies on computer algorithms as opposed to demonstrated properties of mathematical objects. 

In Section 2, we bring together theoretical results our method is based on, and give  a  general formulation of the method.

In Section 3, we concretize the general method
 in the case of Bernoulli observations, providing all the necessary formulas ready for their computer implementation.
 The R source code \citep{R} implementing the calculations in accordance with the formulas can be found in \cite{github}.
 
 Using the computer code, we calculate, for a range of hypothesized values $\theta_0,\theta_1$ and/or error probabilities $\alpha, \beta$, 
 the parameters of optimal sampling plans and their characteristics, as well those of the Wald's sequential probability ratio tests (SPRT) and the fixed-sample-size tests (FSST) with the same levels of type I and type II error probabilities.
 The  results are presented in the form of tables and graphs  and discussed in Section 4.
 

\section{ Kiefer-Weiss problem: general results}

In this section, we formulate the general results on the structure of the optimal tests in the 
 Kiefer-Weiss problem and its modified version.

\subsection{Theoretical basis}
Throughout this paper, we will use the notation of \cite{Novikov2009} and follow its general assumptions. The assumptions are notably more general than we actually need in this paper, but keeping in mind the possible extensions and generalizations, we consider it  convenient to stick to the same framework as in \cite{Novikov2009}.

In particular, we consider randomized sequential tests $\langle\psi,\phi\rangle$, with  $\psi=(\psi_1,\psi_2,$ $\dots)$ being a stopping rule, and $\phi=(\phi_1,\phi_2,\dots)$, being a (terminal) decision rule. It is always assumed that $\psi_n=\psi_n(x_1,\dots,x_n)$ and $\phi_n=\phi_n(x_1,\dots,x_n)$ are measurable functions with values in $[0,1]$, whose values are interpreted as the conditional probabilities, given the data $x_1, \dots,x_n$, to stop at stage $n$, and, respectively, those to reject H$_0$, after stopping has been decided on, for any $n=1,2, \dots$

Denote $c_n^\psi=c_n^\psi(x_1,\dots,x_n)=(1-\psi_1)(1-\psi_2)\dots(1-\psi_{n-1})$, and $s_n^\psi=c_n^\psi\psi_n$, so for any test 
\begin{equation}
\alpha(\psi,\phi)=\sum_{n=1}^{\infty}E_{\theta_0}s_n^\psi\phi_n
\end{equation}
is the type I error probability,  \begin{equation}
\beta(\psi,\phi)=\sum_{n=1}^{
\infty}E_{\theta_1}s_n^\psi(1-\phi_n)
\end{equation}
is the type II error probability, and
\begin{equation}
N(\theta;\psi)=\sum_{n=1}^{\infty}nE_{\theta}s_n^\psi
\end{equation}
is the average sample number when the true parameter value is $\theta$
(provided that $\sum_{n=1}^{\infty}E_{\theta}s_n^\psi=1$, -- otherwise it is infinite).

It is  common that the error probabilities are expressed through the operating characteristic function defined as
 \begin{equation}
OC_\theta(\psi,\phi)=\sum_{n=1}^{
\infty}E_{\theta}s_n^\psi(1-\phi_n):
\end {equation}
$\alpha(\psi,\phi)=1-OC_{\theta_0}(\psi,\phi),\quad\beta(\psi,\phi)=OC_{\theta_1}(\psi,\phi)
$.

Let $\mathscr S(\alpha,\beta)$ be the set of all tests such that
\begin{equation}\label{7_1}
\alpha(\psi,\phi)\leq \alpha,\quad \beta(\psi,\phi)\leq \beta,
\end{equation}
where $\alpha,\beta\in[0,1]$ are some fixed real numbers. 

We are interested in finding the tests that minimize $\sup_\theta N(\theta;\psi)$ over all the tests  in $\mathscr S(\alpha,\beta)$.
This problem is known as the Kiefer-Weiss problem. 

The respective modified Kiefer-Weiss problem is to minimize $N(\theta_*;\psi)$ over all $\langle \psi,\phi\rangle\in \mathscr S(\alpha,\beta)$, for a given fixed value of $\theta_*$.

At the time being, all the known solutions to the Kiefer-Weiss problem, for particular models, are obtained through the modified version of it \citep[see][]{KieferWeiss,Weiss,FreemanWeiss,Lai1973,Lorden,Huffman1983,Zhitlukhin, Tartakovsky2014}.

The solutions to the modified Kiefer-Weiss problem can be obtained, at least in theory,
in a very general situation using the following variant of the Lagrange multipliers method. Let us start with this.

Let \begin{equation}\label{10a_1}L(\psi,\phi)=N(\theta_*;\psi)+\lambda_0\alpha(\psi,\phi)+\lambda_1\beta(\psi,\phi),\end{equation} where $\theta_* $ is some fixed value of the parameter and $\lambda_0,\lambda_1$ are some nonnegative constants (called Lagrange multipliers).

Then the tests minimizing $N(\theta_*;\psi)$ subject to  \eqref{7_1} can be obtained 
through an unconstrained minimization of $L(\psi,\phi)$ over all $\langle\psi,\phi\rangle$,
 using  an appropriate choice of the Lagrange multipliers (see \citep[see][Section 2]{Novikov2009}).

  \cite{Lorden} shows that in the case of i.i.d. observations the problem of minimizing the Lagrangian function is reduced to an optimal stopping problem for a Markov process.

  It is easy to see that finding Bayesian tests used in \cite{KieferWeiss} is mathematically equivalent to the minimization of \eqref{10a_1}.

  To construct the optimal tests we need an additional assumption on the distribution of the observations. Let $f_\theta^n=f_\theta^n(x_1 ,\dots, x_n)$ be the Radon--Nikodym derivative of the distribution of $X_1,\dots,X_n$ with respect to a product-measure $\mu^n=\mu\otimes\dots\otimes \mu$ ($n$ times $\mu$ by itself), $n=1,2,\dots$.

  In \cite{Lorden}, it is shown that, in the case of i.i.d. observations following  a distribution from a Koopman-Darmois family, the tests giving solution to the modified problem have 
 bounded with probability one stopping times when $\theta_*\in(\theta_0,\theta_1)$.

Let us describe the construction of tests minimizing the Lagrangian function calculated at some $\theta_*$,  over all {\em truncated} tests, i.e. those not taking more than a fixed number $H$ of observations ($H$ is also called {\em horizon} in this case).

Formally, let
$\mathscr S^H=\{\langle\psi,\phi\rangle:   \; c_{H+1}^\psi\equiv 0 \}$
be the class of all such tests.
Let us define \begin{equation}\label{14a_1}V_H^H=\min\{\lambda_0f_{\theta_0}^H,\lambda_1f_{\theta_1}^H\},\end{equation} and, recursively over $n=H-1,\dots,1$,
\begin{equation}\label{11a_1}
V_{n}^H=\min\left\{\lambda_0f_{\theta_0}^{n},\lambda_1f_{\theta_1}^{n},
f_{\theta_*}^{n}+\mathcal{I}V^H_{n+1} 
\right\},
\end{equation}
being
\begin{equation}\label{14a_2}
\mathcal{I}V^H_{n+1}=\left(\mathcal{I}V^H_{n+1}\right)(x_1,\dots,x_{n})=\int V^H_{n+1}(x_1,\dots,x_{n+1})d\mu(x_{n+1}).
\end{equation}
\begin{remark}  $V_n^H$ defined above, as well as $L(\psi,\phi)$ in \eqref{10a_1}, implicitly depend on $ \theta_*,\lambda_0,\lambda_1$. 
\end{remark}
From the results of Section 3.1 in \cite{Novikov2009}, we easily obtain the following  characterization of  all the truncated sequential tests minimizing $L(\psi,\phi)$.
\begin{proposition}\label{p1}
For all $\langle\psi,\phi\rangle\in \mathscr S^H$
\begin{equation}\label{11a_2}
L(\psi,\phi)\geq 1+\mathfrak I V_1^H.
\end{equation}
There is an equality in \eqref{11a_2}, for a test
$\langle\psi,\phi\rangle\in \mathscr S^H
$, if and only if it satisfies the following two conditions:
\begin{description}
\item[a)] for all $n=1,2,\dots,H-1$
\begin{equation}\label{11a_3}
I_{\{
\min\{\lambda_0f_{\theta_0}^{n},\lambda_1f_{\theta_1}^{n}\}<
f_{\theta_*}^{n}+\mathcal{I}V^H_{n+1}
\}
}\leq\psi_n\leq I_{
\{ \min\{\lambda_0f_{\theta_0}^{n},\lambda_1f_{\theta_1}^{n}\}\leq
f_{\theta_*}^{n}+\mathcal{I}V^H_{n+1}       \}}
\;
\end{equation}
$\mu^n$-a.e. on $C_n^\psi=\{(x_1,\dots,x_n):c_n^\psi(x_1,\dots,x_n)>0\}$,
and 
\item[b)] for all $n=1,2,\dots,H$
\begin{equation}\label{11a_4}
I_{\{
\lambda_0f_{\theta_0}^n<\lambda_1f_{\theta_1}^n
\}}
\leq\phi_n\leq I_{\{
\lambda_0f_{\theta_0}^n\leq\lambda_1f_{\theta_1}^n
\}}
\end{equation}
$\mu^n$-a.e. on $S_n^\psi=\{(x_1,\dots,x_n):s_n^\psi(x_1,\dots,x_n)>0\}$.
\end{description}
\end{proposition}

Under very mild conditions, the optimal non-truncated tests are obtained on the basis of limits $V_n=\lim_{H\to\infty}V_n^H$. Optimal stopping rules for the non-truncated tests are obtained substituting $V_n$ for $V_n^H$ in \eqref{11a_3} for all $n$, leaving the decision rules  the same, just applying \eqref{11a_4} for all $n$ (see Section 3 in \cite{Novikov2009}).

Let us denote $\mathscr M^H(\theta_*,\lambda_0,\lambda_1)$ the class of all tests satisfying  conditions a) and b)
 of Proposition \ref{p1}, and let $\mathscr M(\theta_*,\lambda_0,\lambda_1)$ be the class of all (non-truncated) tests which satisfy  \eqref{11a_3} with $V_n^H$ is replaced  by $V_n^H$, for all natural $n$,  and satisfy  \eqref{11a_4} for all natural $n$.

The following proposition  shows how the Kiefer-Weiss problem and its modified version can be related (cf. also  Section 3 of \cite{FreemanWeiss}). 
\begin{proposition}\label{p2}
Let $\langle\psi^*,\phi^*\rangle\in\mathscr M(\theta_*,\lambda_0,\lambda_1)$
such that $\alpha=\alpha(\psi^*,\phi^*)$, $\beta=\beta(\psi^*,\phi^*)$, and let
\begin{equation}\label{9a_1}
\sup_\theta N(\theta;\psi^*)-N(\theta_*;\psi^*)=\Delta(\psi^*).
\end{equation}
Then
\begin{equation}\label{13a_1}\sup_\theta N(\theta;\psi^*)\leq \inf\sup_\theta N(\theta;\psi)+\Delta(\psi^*),\end {equation}
where the infimum is taken over all ${\langle\psi,\phi\rangle\in \mathscr S(\alpha,\beta)}.
$
\end{proposition}
{\bf Proof.}
Let $\langle\psi^*,\phi^*\rangle$ be a test satisfying the conditions of Proposition \ref{p2}, and let $\langle\psi,\phi\rangle$ be any test from $\mathscr S(\alpha,\beta)$.
Then
$$
N(\theta_*;\psi^*)+\lambda_0\alpha+\lambda_1\beta=
N(\theta_*;\psi^*)+\lambda_0\alpha(\psi^*,\phi^*)+\lambda_1\beta(\psi^*,\phi^*)$$
$$
\leq 
N(\theta_*;\psi)+\lambda_0\alpha(\psi,\phi)+\lambda_1\beta(\psi,\phi)
\leq N(\theta_*;\psi)+\lambda_0\alpha+\lambda_1\beta, 
$$
where the first inequality is due  to   Proposition \ref{p1}.
Therefore, $N(\theta_*;\psi^*)\leq N(\theta_*;\psi)$, so 
$$
\sup_\theta N(\theta;\psi^*)=N(\theta_*;\psi^*)+\Delta(\psi^*)\leq N(\theta_*;\psi)+\Delta(\psi^*)\leq \sup_\theta N(\theta;\psi)+\Delta(\psi^*),
$$
and \eqref{13a_1} follows.
$\Box$

Obviously, if $\Delta(\psi^*)=0$ in Proposition \ref{p2}, then $\langle \psi^*,\phi^*\rangle$ solves the original Kiefer-Weiss problem with  $\alpha=\alpha(\psi^*,\phi^*)$ and $\beta=\beta(\psi^*,\phi^*)$.
In this way, Proposition \ref{p2} reduces the original Kiefer-Weiss problem to its modified variant: one has to seek for a solution to the modified Kiefer-Weiss problem, with some $\theta_*$, for which $N(\theta_*;\psi^*)$ is a maximum of $N(\theta;\psi^*)$ over all $\theta$.

The treatment of the Kiefer-Weiss problem by \cite{Weiss}, in particular symmetric cases of sampling from  normal and Bernoulli distributions, in essence, makes use of this proposition.

 \cite{FreemanWeiss} propose, for the non-symmetric Bernoulli case of  observations, a choice of $\theta_*$  making $\Delta(\psi^*)$ relatively small which gives nearly optimal tests in the Kiefer-Weiss problem.
\subsection{The proposed computational method}
The method  is based on Propositions \ref{p1} and \ref{p2}.

Proposition \ref{p1} is used for the Lagrangian minimization when seeking for  solutions to the modified problem potentially solving the Kiefer-Weiss problem by virtue of  Proposition \ref{p2}. 

Both parts are essentially numerical, and we will show in the next Section, in the case  of sampling  from a Bernoulli population,  how  they can be implemented using the modern statistical software.

In what remains of this Section we want to give a general description of the proposed method.

The method deals with an appropriate choice of the constants $\theta_*,\lambda_0,\lambda_1 $. From any class $\mathscr M^H(\theta_*,\lambda_0,\lambda_1)$ let us choose one test (for example, the easiest for implementation), - in this way,  $\mathscr M^H(\theta_*,\lambda_0,\lambda_1)$ will identify a specific test.

We assume that a computer procedure is available which calculates, for any $\mathscr M^H(\theta_*,\lambda_0,\lambda_1)$, the whole set of test characteristics: the error probabilities and the average sample number, whatever be the true value of $\theta$. Derived from these, 
we will also assume that computing $\Delta(\psi^*)$ (see \eqref{9a_1}) is  available as well.

We know from \cite{Lorden} that in the case of one-parametric Koopman-Darmois family of i.i.d. observations there is an upper bound on the horizon $ H $ of the optimal test in the modified Kiefer-Weiss problem when $\theta_*$ is between $\theta _0$ and $\theta_1$. Generally, it depends on $\theta_0,\theta_1, \theta_*,\lambda_0  $ and $\lambda_1$ (see \cite{Lorden}). Because $\theta_0$ and $\theta_1$ will not be moved within the algorithm, let us only retain $\theta_*,\lambda_0,\lambda_1$ in the notation, supposing  $H=H(\theta_*,\lambda_0,\lambda_1)$ is  available for computation in such case.

Obviously, there can be cases where the bound does not exist, or is not available for computing, for some reason, or we just do  not want to use it. We want the method be applicable in both cases, more precisely, we will treat these cases as two variants of the method because they are based on the same principles.

{\flushleft\bf The proposed method }\\
{\bf Option 1.} Supposing the bound $H=H(\theta_*,\lambda_0,\lambda_1)$ is available.
\begin{enumerate}
\item Start with some $\lambda_0,\lambda_1$.
\item
 Seek for a $\theta_*$ such that the test $\langle\psi^*,\phi^*\rangle\in\mathscr M^H(\theta_*,\lambda_0,\lambda_1))$ with $H=H(\theta_*,\lambda_0,\lambda_1)$ has  a minimum value of $\Delta(\psi^*)$ over all $\theta$:
\begin{equation}\Delta(\psi^*)\leq\Delta(\psi)\quad\mbox{for}\;\langle\psi,\phi\rangle\in\mathscr M^H(\theta,\lambda_0,\lambda_1)\end{equation} 
with $H=H(\theta,\lambda_0,\lambda_1)$, whatever $\theta$.
\item
Evaluate $\alpha=\alpha(\psi^*,\phi^*)$ and $\beta=\beta(\psi^*,\phi^*)$. If they do not comply with requirements on the error probabilities, repeat steps  2 and 3 with other $\lambda_0$ and $\lambda_1$.
\end{enumerate}
If $\Delta(\psi^*)=0$ (because of the computational nature of the method, this should mean in fact that it is close to 0), the solution  to the Kiefer-Weiss problem $\langle\psi^*,\phi^*\rangle$ for given $\alpha=\alpha(\psi^*,\phi^*)$ and $\beta=\beta(\psi^*,\phi^*)$  is found. If $\Delta(\psi^*)$ is not very close to 0, it could be informative anyway, because it shows how good the modified version is  for the Kiefer-Weiss problem.

 We implement this method in the next Section for the general  Bernoulli case and obtain, for a series of particular hypothesized values of parameters and probability errors, numerical results showing that the minimum value of $\Delta(\psi^*)$ is 0 in each case. 

 In this way, we may speak about numerical solution of the Kiefer-Weiss problem for the Bernoulli observations.
 We provide full computational algorithms which make our results completely  verifiable.
 
At the same time, we are rather pessimistic about the possibility of a strictly mathematical proof that this will always be the case, even for Koopman-Darmois families, and even for the Bernoulli case.

{\bf Option 2.} Not using bounds for the maximum sample size.
\begin{enumerate}
\item Start with some $\lambda_0,\lambda_1$.
\item For an increasing sequence of $H$ repeat:
\item
 Seek for a $\theta_*$ such that the test $\langle\psi^*,\phi^*\rangle\in\mathscr M^H(\theta_*,\lambda_0,\lambda_1))$  has  a minimum value of $\Delta(\psi^*)$ over all $\theta$:
\begin{equation}\Delta(\psi^*)\leq\Delta(\psi)\quad\mbox{for}\;\langle\psi,\phi\rangle\in\mathscr M^H(\theta,\lambda_0,\lambda_1),\end{equation} 
 whatever $\theta$. Evaluate  $L(\psi^*,\phi^*)$.
\item Stop repeating when the successive values of $L_H(\psi^*,\phi^*)$ are close to each other.
\item
Evaluate $\alpha=\alpha(\psi^*,\phi^*)$ and $\beta=\beta(\psi^*,\phi^*)$. If they do not comply with requirements on the error probabilities, repeat steps 2 through 5 with other $\lambda_0$ and $\lambda_1$.
\end{enumerate}

 Option 2, for any given horizon $H $ tries to find a solution to the Kiefer-Weiss problem in the class of all truncated, at level $H$, sequential tests. If the solution is a truncated test, it will be found when $H$ is sufficiently large. If the optimal test is not truncated, the algorithm tries to find a good approximation to it in the class of the  truncated tests, with high levels of truncation, which may be considered a numerical solution of the Kiefer-Weiss problem, at some precision level. 

 To see the particular usefulness of this approach, one can have a look at the example of the uniform distribution  \citep[see Section 3.1 of ][starting from  p. 148]{NovikovPalacios} . 
 It is easily seen that in this case the algorithm applies exactly (not numerically), and that for any fixed $H$ it readily gives an optimal truncated test, whose characteristics  converge, as $H\to\infty$, to those of the optimal non-truncated test, solving the Kiefer-Weiss problem for this particular model. Interestingly, the solution to the Kiefer-Weiss problem is an SPRT in this case. 

The virtue of this example is more theoretical than practical, though. 

There are other examples (rather artificial as well) of  the modified Kiefer-Weiss problem, where the optimal stopping rule has a non-bounded stopping time \citep{Hawix}, but there is no Kiefer-Weiss problem it is related to. Should a general context for this example exist, Option 2 should be applicable for its numerical solution. We are pretty sure that the numerical optimization in step 3 can be available for its computer implementation much more generally than only in the uniform and the Bernoulli case.

Option 1 is preferable (when applicable), because it avoids the iteration cycle over truncation level $H$ in Option 2. But Option 2 is more general and is applicable to virtually any Kiefer-Weiss problem for which  the computation in step 3 is feasible. Numerical experiments in the example case of the next Section, where both Options are applicable, show that both give the same results.

\section{Kiefer-Weiss problem for sampling from a  Bernoulli population   }

In this section, we obtain  formulas for solving the modified Kiefer-Weiss problem and implement the computational algorithms of the preceding Section for solving the Kiefer-Weiss problem in  the particular case of sampling from  a Bernoulli population. On the basis of this, we obtain and analyze numerical results for the efficiency of the numerical solution to the Kiefer-Weiss problem with respect to the sequential probability ratio test and to the fixed-sample-size test, ones with the same error probabilities.

Let the observations $X_1,X_2,\dots,X_n,\dots$ be independent identically distributed Bernoulli random variables with $f_\theta(x)=\theta^x(1-\theta)^{1-x}$, for $x=0,1$, and $0<\theta<1$.
Then the joint probability $$f_\theta^n(x_1,\dots,x_n)=g_\theta^n(s_n)=\theta^{s_n}(1-\theta)^{n-s_n},$$ where $s_n=\sum_{i=1}^n x_i$, $n=1,2,\dots$
 
\subsection{Construction of tests}\label{s14a_1}
Let  $0<\theta_0<\theta_1<1$ be two fixed hypothesized parameter values.

Let us construct, using Proposition \ref{p1}, a solution to the modified Kiefer-Weiss problem for a given $\theta_*$ (see \eqref{14a_1} -- \eqref{14a_2}).

In this Bernoulli model, it is not difficult to see, by induction, that 
$$V_n^H(x_1,\dots,x_n)=U_n^H\left(\sum_{i=1}^nx_i\right),$$ for  $n=1,\dots, H$, where
\begin{equation}\label{19a_1}U_H^H(s)=\min\{\lambda_0g_{\theta_0}^H(s),\lambda_1g_{\theta_1}^H(s)\},\;s=0,1,\dots, H\end{equation} and, recursively over $n=H-1,\dots, 1$,
\begin{equation}\label{19a_2}
U_n^H(s)=\min\{\lambda_0g_{\theta_0}^n(s),\lambda_1g_{\theta_1}^n(s),
g_{\theta_*}^n(s)+ U_{n+1}^H(s+1)+U_{n+1}^H(s)\},\end{equation}
$s=0,1,\dots, n$.

Let us consider  a non-randomized test $\langle\psi^*,\phi^*\rangle\in\mathscr M^H(\theta_*,\lambda_0,\lambda_1)$, which, 
at any stage $n\leq H-1$, with $s_n=\sum_{i=1}^n x_i$ observed,
\begin{itemize}
\item [(a)]stops and accepts H$_0$, if $\lambda_1g_{\theta_1}^n(s_n)= U_n^H(s_n)$ (in which case $\psi_n^*=1,\;\phi_n^*=0$),
\item[(b)] stops and rejects H$_0$, if $\lambda_0g_{\theta_0}^n(s_n)= U_n^H(s_n)$
(giving preference to (a) if both (a) and (b)  apply), being in this case $\psi_n^*=1,\;\phi_n^*=1$, and
\item [(c)]continues to the next stage, if neither (a) nor (b) applies (being $\psi_n^*=0$ in this case);
\end{itemize}
and, at stage $H$, stops and accepts H$_0$ if $\lambda_0g_{\theta_0}^H(s_H)\geq \lambda_1g_{\theta_1}^H(s_H)$ 
($\phi_H^*=0$) and rejects H$_0$ otherwise ($\phi_H^*=1$).

\subsection{Operating characteristic, average sample number and related formulas}

Let \begin{equation}\label{25a_1}a_\theta^H(s)=g_\theta^H(s)(1-\phi_H^*(s)),\; s=0,1,\dots,H,\end{equation} and, recursively over $n=H-1,H-2,\dots,1$, 
\begin{equation}\label{25a_2}a_\theta^{n}(s)=g_\theta^{n}(s)\psi_{n}^*(s)(1-\phi_n^*(s))+a_\theta^{n+1}(s)+a_\theta^{n+1}(s+1),\;s=0,1,\dots,n.\end{equation}
Then the error probability of type II is  $\beta(\psi^*,\phi^*)=a_{\theta_1}^0=a_{\theta_1}^1(0)+a_{\theta_1}^1(1)$.

To understand this, one can trace the appearance, in \eqref{19a_1}--\eqref{19a_2}, of all the terms containing $\lambda_1$. The terms having  $\lambda_1$ as a coefficient are exactly those in \eqref{25a_1}--\eqref{25a_2} (with $\theta=\theta_1$). Now,  take into account that, by Proposition \ref{p1}, $U_0^H=1+U_1^H(0)+U_1^H(1)$ 
coincides with $$\lambda_0\alpha(\psi^*,\phi^*)+\lambda_1\beta(\psi^*,\phi^*)+N(\theta_*;\psi^*),$$
and the term in $U_0^H$ having $\lambda_1$ as a coefficient is $a_{\theta_1}^0$, so it is equal to $\beta(\psi^*,\phi^*)$.

Because now $\beta(\psi^*,\phi^*)=OC_{\theta_1}(\psi^*,\phi^*)=a_{\theta_1}^0$,
we conclude that $OC_{\theta}(\psi^*,\phi^*)=a_{\theta}^0$ for any $\theta$. And finally
$\alpha(\psi^*,\phi^*)=1-OC_{\theta_0}(\psi^*,\phi^*)$.

In a similar way, let \begin{equation}\label{26a_1}
                       b_\theta^H(s)=0,\; s=0,1,\dots,H,
                      \end{equation}
 and, recursively over $n=H-1,H-2,\dots,1$, 
\begin{equation}\label{26a_2}
b_\theta^{n}(s)=(g_\theta^{n}(s)+b_\theta^{n+1}(s)+b_\theta^{n+1}(s+1))(1-\psi_{n}^*(s)), \; s=0,1,\dots,n.
\end{equation}
Then the average sample number $N(\theta;\psi^*)=1+b_\theta^0$, where $b_\theta^0=b_\theta^1(0)+b_\theta^1(1)$.

When high levels of truncation $H$ are needed, the following variant of \eqref{25a_1}--\eqref{25a_2} seems to work computationally better.

Let us denote $G^n_\theta(s)={n\choose s}\theta^s(1-\theta)^{n-s}={n\choose s}g_\theta^n(s)$, $s=0,1,\dots,n$.

Then, define
 \begin{equation}\label{26a_3}A_\theta^H(s)=G_\theta^H(s)(1-\phi_H^*(s)),\; s=0,1,\dots,H,\end{equation} and, recursively over $n=H-1,H-2,\dots,1$, 
\begin{equation}\label{26a_4}A_\theta^{n}(s)=G_\theta^{n}(s)\psi_{n}^*(s)(1-\phi_n^*(s))+A_\theta^{n+1}(s)\frac{n+1-s}{n+1}+A_\theta^{n+1}(s+1)\frac{s+1}{n+1},\end{equation}
$s=0,1,\dots,n$.

It is easy to see, by induction, that $A_\theta^n(s)={n\choose s} a_\theta^n(s)$, $s=0,1,\dots, n$, $n=1,2,\dots, H$, so
\begin{equation}\label{30a_2}OC_{\theta}(\psi^*,\phi^*)=A_{\theta}^0=A_\theta^1(0)+A_\theta^1(1).\end{equation}

Analogously, let  \begin{equation}\label{26a_5}
                       B_\theta^H(s)=0,\; s=0,1,\dots,H,
                      \end{equation}
 and, recursively over $n=H-1,H-2,\dots,1$, 
\begin{equation}\label{26a_6}
B_\theta^{n}(s)=(G_\theta^{n}(s)+B_\theta^{n+1}(s)\frac{n+1-s}{n+1}+B_\theta^{n+1}(s+1)\frac{s+1}{n+1})(1-\psi_{n}^*(s)), 
\end{equation}
$ s=0,1,\dots,n.$

Again, $B_\theta^n(s)={n\choose s} b_\theta^n(s)$, $s=0,1,\dots, n$, $n=1,2,\dots, H$,
and the average sample number \begin{equation}\label{30a_1}N(\theta;\psi^*)=1+B_\theta^0,\; \mbox{where} \;B_\theta^0=B_\theta^1(0)+B_\theta^1(1).\end{equation}

We implement, in the R program code, versions  \eqref{26a_3}-\eqref{26a_4}  and \eqref{26a_5}-\eqref{26a_6}, repectively,  as a routine for calculation of the operationg characteristic function \eqref{30a_2} and the average samle number \eqref{30a_1}. 

It is very likely that the algorithms above in this subsection are applicable not only to the tests obtained through the Lagrange minimization but also generally to any truncated test. We defer the strict proof of this fact to  a later occasion.

We will use  Option 1 of the method  in  Section 2, in view of the intensity of calculations we need 
for obtaining the massive numerical results for the analysis below in this Section. Option 2 gives the same results but is somewhat slower.

Due to \cite{Lorden} (see also \cite{Hawix}) 
the test minimizing the Lagrangian function, given $\theta_*,\lambda_0, \lambda_1$, can be found in 
$\mathscr M^H(\theta_*,\lambda_0,\lambda_1)$ with  $H$  not exceeding
\begin{equation}\label{30a_3}
H(\theta_*,\lambda_0,\lambda_1)=\inf\{n\geq 1:a\log \lambda _0+b\log \lambda_1-n\leq (a+b) \log w_0\},
\end {equation}
where $w_0=\left(1-P_{\theta_0 } (f_{\theta_0}(X)<f_{\theta_1}(X))-P_{\theta_1 } (f_{\theta_0}(X)\geq f_{\theta_1}(X))\right)^{-1}
$, and $a$ and $b$ are determined from
$$
a\log \left(\frac{ f_{\theta_*}(X)}{f_{\theta_0}(X))}\right)+b\log\left(\frac{ f_{\theta_*}(X)}{f_{\theta_1}(X))}\right)\equiv 1.
$$

Thus, the computer implementation of \eqref{30a_3} is straightforward.

At last, for the minimization step at stage 2 of the Option 1 of the method in preceding Section, we use  R's {\tt optimize} function, first, to find the maximum of $N(\theta:\psi^*)$ over $\theta\in(\theta_0,\theta_1)$,  given $\theta_*$, then the minimum of $\sup_\theta N(\theta;\psi^*)-N(\theta_*:\psi^*)$, over all $\theta_*  \in(\theta_0,\theta_1)$. We use the default parameter of tolerance when using {\tt optimize}, which  is approx. 0.00012 in our case.
Using lower levels of the tolerance parameter, better approximation is achieved.

The details of the implementation can be consulted in \cite{github}.  
 
\subsection{Numerical results}
 
The main goal of this part is to illustrate the use of the developed R code on concrete examples based on the Bernoulli sampling,  to show that the obtained sequential tests in any one of the examples provide numerical solutions to the Kiefer-Weiss problem, and to analyze the efficiency of the obtained tests with respect to the classical sequential probability ratio tests and the fixed-sample-size tests provided these have the same level of error probabilities.

We use a series of examples seen in \cite{FreemanWeiss}, specified by 5 pairs of hypothesized success probabilibies: $\theta_0=0.05,\theta_1=0.15$; $\theta_0=0.1,\theta_1=0.2$;  $\theta_0=0.2,\theta_1=0.3$; $\theta_0=0.4,\theta_1=0.5$ and
$\theta_0=0.45,\theta_1=0.55$. In each one, we employ a range of error probabilities commonly used in practice: $\alpha(=\beta)=$ 0.1, 0.05, 0.025, 0.01, 0.005, 0.001 and 0.0005.

For each combination of $\theta_0,\theta_1$ and $\alpha=\beta$ we ran the computer code corresponding to the implementation of the method of Section 2 (Option 1), seeking for a test with the  closest values of the error probabilities to their nominal values. In each case the real and the nominal error probability are within 0.001 of relative distance to each other.

The respective results are presented in Tables 1  -- 5. For each test, there are its corresponding values of $\theta_*,\lambda_0,\lambda_1$ in the table, as well as its average sample number $N(\theta_*;\psi^*)$   and its corresponding $\Delta(\psi^*)$, and the 0.99-quantile Q$_{.99}(\psi^*) $ of the distribution of the sample number, under $\theta_*$ as well. We also  present the maximum sample number (denoted $H$ in the table) the test actually takes (this is not the upper bound \eqref{30a_3}).

In the second part of each table, there are the calculated characteristics of the corresponding SPRT with the closest values of $\alpha$ and $\beta$ to the nominal ones.
Those are the values of the average sample number $N(\theta_*;W)$ and the 0.99-quantile Q$_{.99}(W)$ of the  distribution of the sample number, both calculated at $\theta_*$. They are calculated  using the exact  formulas 
in \cite{Young}, and not through the Wald approximations, as in  \cite{FreemanWeiss}.
$\log A $ and $\log B$ are the endpoints of the continuation interval of the corresponding SPRT.

At last, FSS is the minimum value  of the sample number required by the optimal  fixed-sample-size test with error probabilities not exceeding $\alpha$ and $\beta$. It is calculated using the binomial distribution of the test statistic rather than its normal approximation used in \cite{FreemanWeiss}.

In the last part of each table, there are calculated values of efficiency of each test with respect to the FSST. The efficiency is calculated as the ratio of FSS to other characteristics of the respective test: : $R(\psi^*)={FSS}/{N(\theta_*;\psi^*)}$ and $QR(\psi^*)={FSS}/{Q_{.99}(\psi^*)}$ for the optimal Kiefer-Weiss test and $R(W)={FSS}/{N(\theta_*;W)}$ and $QR(W)={FSS}/{Q_{.99}(W)}$ for the Wald's SPRT. For example, $R(\psi^*)=1.5$ means the optimal Kiefer-Weiss test takes   1.5 times fewer observations, on the average, than the corresponding fixed-sample-size test.

There is no SPRT part in Table 5. The reason for this is that in the symmetric case (when  $\theta_0=1-\theta_1$), unlike other cases, it is generally impossible to find an SPRT  matching, at least approximately, the nominal values of $\alpha$ and $\beta$. This is because,  in the symmetric case,  there is a restricted number of available values of $\alpha=\beta$, for example, for $\theta_0=0.45$ and $\theta_1=0.55$, these are 0.0991, 0.0826, 0.0686, 0.0568, 0.0470, etc., none of them exactly matching the values of $\alpha=\beta$ used for other hypothesis  pairs.
 In the symmetric case, the SPRT's decision-making process is equivalent to a random walk within  two bounds formed by two horizontal straight lines, so the values of  $\alpha=\beta$ we gave above were obtained from the well-known formula for  Gambler's Ruin probability when the lines are symmetric with respect to the horizontal axis.

To evaluate the performance of the optimal tests for $\alpha\not=\beta$ we ran the R computer code on a grid of equidistant points of $\lambda_0$ and $\lambda_1$ on the logarithmic scale, for each pair of $\theta_0$ and $\theta_1$ finding, at each point of the grid, the optimal test $\langle\psi^*,\phi^*\rangle$ for the Kiefer-Weiss problem. In each case $\Delta(\psi^*)$ was found to be 0 (within the given precision of  calculations). 

 For each pair of $\lambda_0$ and $\lambda_1$, we calculated the error probabilities $\alpha=\alpha(\psi^*,\phi^*)$ and $\beta=\beta(\psi^*,\phi^*)$ and the average sample numbers $N(\theta_*;\psi^*)$, $N(\theta_0;\psi^*)$  and $N(\theta_1;\psi^*)$ corresponding to the optimal test $\langle\psi^*,\phi^*\rangle\in\mathscr M(\theta_*,\lambda_0,\lambda_1)$, as well as an estimate of the fixed sample size required  to achieve the same $\alpha $ and $\beta$.
This time we use an approximate formula for the FSS based on the normal approximation
$$
FSS\approx \left(\frac{z_\alpha \sqrt{\theta_0(1-\theta_0)}+z_\beta \sqrt{\theta_1(1-\theta_1)}}{\theta_1-\theta_0}\right)^2,
$$
where $z_\alpha=\Phi^{-1}(1-\alpha)$, being $\Phi$ the cumulative distribution function of the standard normal distribution. This form is preferable from the point of view of   smoothness of the graphical representation below, while the relative error, in comparison with its exact value based on the binomial distribution is within as much as 5\%, for the range of the   $\alpha$ and $\beta$ calculated. 

In Figures 1 -- 3, we present the results of the performance evaluation for three hypothesis pairs: $\theta_0=0.05$ vs. $\theta_1=0.15$, $\theta_0=0.2$ vs. $\theta_1=0.3$  and $\theta_0=0.45$ vs. $\theta_1=0.55$.

Each graph depicts one performance characteristic ($z$-coordinate) as a function of $\alpha$ ($x$-coordinate) and $\beta$ ($y$-coordinate). For $\alpha $ and $\beta$, the scale of decimal logarithms is used. In each Figure, the upper left graph represents the same efficiency  used in the Tables, defined as $R(\psi^*)=FSS/N(\theta_*;\psi^*)$. The upper right graph depicts the average sample number $N(\theta_*;\psi^*)$ as a function of $\alpha$ and $\beta$. 
The two lower graphs in each Figure represent the performance of the optimal tests under $H_0$ and $H_1$ (very much analogously to the first graph) depicting the efficiency defined as $R_0(\psi^*)=FSS/N(\theta_0;\psi^*)$ and $R_1(\psi^*)=FSS/N(\theta_1;\psi^*)$, respectively.

The graphs are prepared using the 3D visualization  package {\tt rgl} \citep{rgl}.
Each graph is based on a grid of 25$\times$25 equidistant points of $\ln\lambda_0$ and  $\ln\lambda_1$, both within a range of 6 to 13. 

There are interactive versions of the graphs in \cite{github}.

\begin{table}[!p]
\begin{tabular}{rrrrrrrr}
$\alpha=\beta$&0.1&0.05&0.025&0.01&0.005&0.001&0.0005
\\
\hline
$\theta_*$&0.0768	&0.0823	&0.0848&	0.0865&	0.0874&	0.0885&	0.0888
\\
$\lambda_0$&157.70	&356.55&	785.75&	2067.24	&4334.60&	22957.0&	46319.1
\\
$\lambda_1$&193.35&	430.27&	952.99	&2512.30	&5183.48&	27541.8&	56073.6
\\
H&128&	182&	237	&302&	345&	465&	508
\\$N(\theta_*;\psi^*)$&38.62&	64.75&	93.79&	134.33&	167.10&	246.23	&281.64
\\$\Delta(\psi^*)$&-4.E-8&	1.E-4&	9E-5	&4.E-5	&4.E-5&	-3.E-6&	-4.E-5
\\
$Q_{.99}(\psi^*)$&89	&134&	187&	244&	291&	397&	442
\\
\hline
$\log A$&-2.13&	-2.90&	-3.60	&-4.54	&-5.24&	-6.85&	-7.55
\\$\log B$&1.85&	2.59&	3.27&	4.22&	4.91	&6.53	&7.22
\\$N(\theta_*;W)$&40.53&	71.93&	109.81&	173.09&	228.77	&388.01&	469.32
\\$Q_{.99}(W)$&149	&269&	414&	660&	884	&1502&	1819
\\
\hline
FSS&60	&93&	136&	181&	224&	322&	365
\\
\hline
R$(\psi^*)$&1.55&	1.44&	1.45&	1.35&	1.34	&1.31&	1.30
\\
QR$(\psi^*)$&0.67&	0.69&	0.73&	0.74&	0.77&	0.81&	0.83
\\R$(W)$&1.48&	1.29&	1.24&	1.05&	0.98&	0.83&	0.78
\\
QR$(W)$&0.40&	0.35&	0.33	&0.27&	0.25&	0.21&	0.20
\\
\end{tabular}
\caption{Optimal tests for $\theta_0=0.05$, $\theta_1=0.15$ }
\end{table}
\begin{table}[!p]
\begin{tabular}{rrrrrrrr}
$\alpha=\beta$&0.1&0.05&0.025&0.01&0.005&0.001&0.0005
\\ 
\hline
$\theta_*$&0.1364&0.1394&0.1409&0.1420&0.1425&0.1432&0.1434\\
$\lambda_0$&246.64&557.46&1204.68&3212.66&6701.95&35404.1&72301.3
\\
$\lambda_1$&275.38&620.37&1343.38&3595.28&7471.57&39608.9&80459.6
\\
H&232&308&384&480&549&721&783
\\
$N(\theta_*;\psi^*)$&56.45&95.03&138.06&198.54&246.92&364.39&416.74
\\$\Delta(\psi^*)$&5.E-5&9.E-7&9.E-5&5.E-7&2.E-5&2.E-5&4.E-5
\\
$Q_{.99}(\psi^*)$&135&205&274&364&434&592&657
\\
\hline
$\log A$&-2.13&-2.88&-3.60&-4.53&-5.24&-6.85&-7.54
\\$\log B$&1.95&2.70&3.42&4.35&5.04&6.66&7.35\\
$N(\theta_*;W)$& 59.48&106.66&163.68&258.59&342.82&583.22&705.82\\
$Q_{.99}(W)$&224&410&636&1007&1337&2280&2759\\
\hline
FSS& 86&135&190&272&328&479&541\\
\hline
R$(\psi^*)$& 1.52&1.42&1.38&1.37&1.33&1.31&1.30\\
QR$(\psi^*)$& 0.64&0.66&0.69&0.75&0.76&0.81&0.82\\
R$(W)$&1.45&1.27&1.15&1.05&0.96&0.82&0.77\\
QR$(W)$& 0.38&0.33&0.30&0.27&0.25&0.21&0.20\\
\end{tabular}
\caption{Optimal tests for $\theta_0=0.1$, $\theta_1=0.2$ }
\end{table}
\begin{table}[!p]
\begin{tabular}{rrrrrrrr}
$\alpha=\beta$&0.1&0.05&0.025&0.01&0.005&0.001&0.0005\\
\hline
$\theta_*$&0.2435&0.2450&0.2457&0.2462&0.2464&0.2468&0.2469
\\
$\lambda_0$&381.04&866.23&1870.27&4985.94&10346.1&54837.5&111392
\\
$\lambda_1$&403.11&912.57&1971.23&5256.63&10880.4&57566.7&117078
\\
H&414&539&653&786&895&1133&1242
\\$N(\theta_*;\psi^*)$&84.43&142.50&206.73&297.74&370.24&546.67&625.28
\\$\Delta(\psi^*)$&2.E-5&1.E-5&1.E-8&1.E-7&2.E-5&1.E-5&1.E-5
\\
$Q_{.99}(\psi^*)$&204&309&410&547&649&886&989
\\
\hline
$\log A$&-2.12&-2.88&-3.59&-4.53&-5.23&-6.84&-7.54
\\$\log B$&2.05&2.78&3.51&4.44&5.14&6.75&7.44
\\$N(\theta_*;W)$&89.41&160.30&247.69&389.73&517.48&880.57&1066.27
\\$Q_{.99}(W)$&346&627&968&1527&2027&3453&4181
\\
\hline
FSS&127&204&289&402&495&713&806
\\
\hline
R$(\psi^*)$&1.50&1.43&1.40&1.35&1.34&1.30&1.29
\\
QR$(\psi^*)$&0.62&0.66&0.70&0.73&0.76&0.80&0.82
\\
R$(W)$&1.42&1.27&1.17&1.03&0.96&0.81&0.76
\\
QR$(W)$&0.37&0.33&0.30&0.26&0.24&0.21&0.19
\\
\end{tabular}
\caption{Optimal tests for $\theta_0=0.2$, $\theta_1=0.3$ }
\end{table}
\begin{table}[!p]
\begin{tabular}{rrrrrrrr}
$\alpha=\beta$&0.1&0.05&0.025&0.01&0.005&0.001&0.0005\\
\hline
$\theta_*$&0.4490&	0.4493&	0.4494	&0.4494&	0.4495&	0.4495&	0.4495
\\
$\lambda_0$&519.98	&1177.67	&2540.72 &	6804.60&	14096.5&	74239.7&	151569
\\
$\lambda_1$&524.39&	1186.52&	2561.23&	6853.52&	14170.5&	74851.8&	152375\\
H&575&	744&	893&	1082&	1231&	1558&	1698
\\$N(\theta_*;\psi^*)$&	111.82&	189.19&	274.56&	395.75&	491.95	&726.33&	830.78\\
$\Delta(\psi^*)$&-6.E-6&	-2.E-5&	6.E-7&	3.E-5	&8.E-5&	1.E-4&	8.E-5
\\
$Q_{.99}(\psi^*)$&272&	409&	546&	726&	862&	1177&	1313
\\
\hline
$\log A$&		-2.12&	-2.86	&-3.58&	-4.51&	-5.21&	-6.82&	-7.52
\\$\log B$&2.10&	2.85&	3.57&	4.50&	5.20&	6.81&	7.51
\\$N(\theta_*;W)$&118.79&	213.16&	330.10&	519.05&	688.86&	1172.59&	1420.01\\
$Q_{.99}(W)$&465&	835&	1293&	2037&	2703&	4605&	5576\\
\hline
FSS&	168&	268&	384&	535	&655&	944	&1071
\\
\hline
R$(\psi^*)$&1.50&	1.42&	1.40&	1.35&	1.33&	1.30&	1.29
\\
QR$(\psi^*)$&0.62&	0.66&	0.70&	0.74&	0.76&	0.80&	0.82\\
R$(W)$&1.41	&1.26&	1.16&	1.03&	0.95&	0.81&	0.75
\\
QR$(W)$&0.36	&0.32&	0.30&	0.26&	0.24&	0.21&	0.19
\\
\end{tabular}
\caption{Optimal tests for $\theta_0=0.4$, $\theta_1=0.5$ }
\end{table}
\begin{table}[!p]
\begin{tabular}{rrrrrrrr}
$\alpha=\beta$&0.1&0.05&0.025&0.01&0.005&0.001&0.0005\\\hline
$\theta_*$&0.5000&0.5000&0.5000&0.5000&0.5000&0.5000&0.5000\\
$\lambda_0$&526.61&1193.78&2577.11&6878.97&14273.4&75384.0&153475\\
$\lambda_1$&526.61&1193.78&2577.11&6878.97&14273.4&75383.2&153475\\
H&571&733&887&1081&1227&1557&1699
\\$N(\theta_*;\psi^*)$&112.71&191.27&277.26&399.83&497.03&733.80&839.32
\\$\Delta(\psi^*)$&0.E+0&0.E+0&6.E-14&-1.E-13&-3.E-13&2.E-13&2.E-13
\\
$Q_{.99}(\psi^*)$&274&414&551&733&871&1190&1328\\
\hline
FSS&163&269&383&539&661&951&1077
\\
\hline
R$(\psi^*)$&1.45&1.41&1.38&1.35&1.33&1.30&1.28\\
QR$(\psi^*)$&0.59&0.65&0.70&0.74&0.76&0.80&0.81
\end{tabular}
\caption{Optimal tests for $\theta_0=0.45$, $\theta_1=0.55$ }
\end{table}
 \begin{figure}
\includegraphics[scale=0.5]{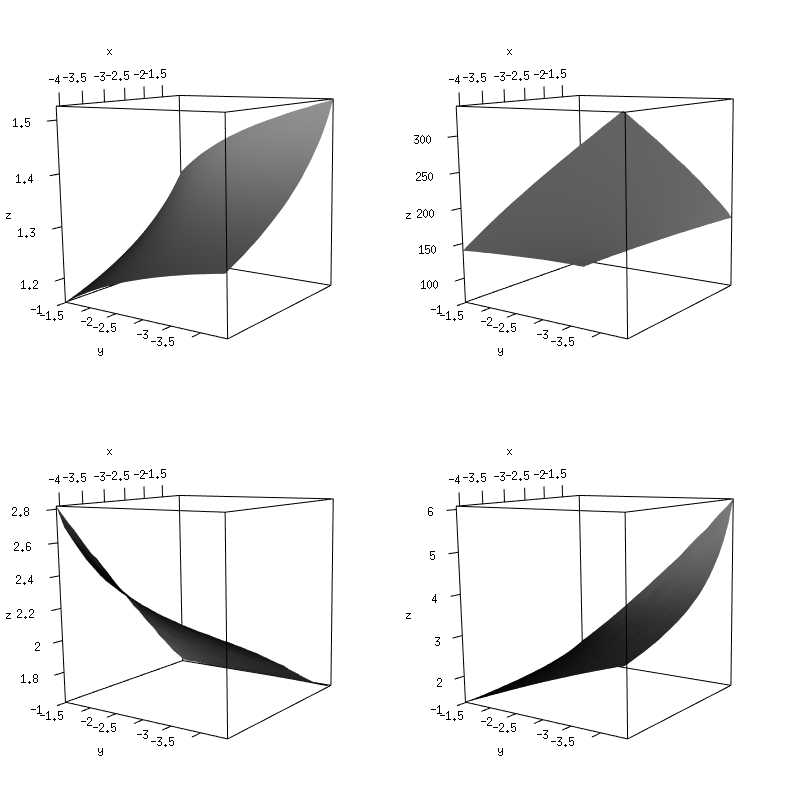}
 \caption{Performance of the optimal test for $\theta_0=0.05$ vs.  $\theta_1=0.15$:\\ $R(\psi^*)$ and $N(\theta_*;\psi^*)$ (above), $R_0(\psi^*)$ and $R_1(\psi^*)$ (below)}

 \end{figure}
\begin{figure}
\includegraphics[scale=0.5]{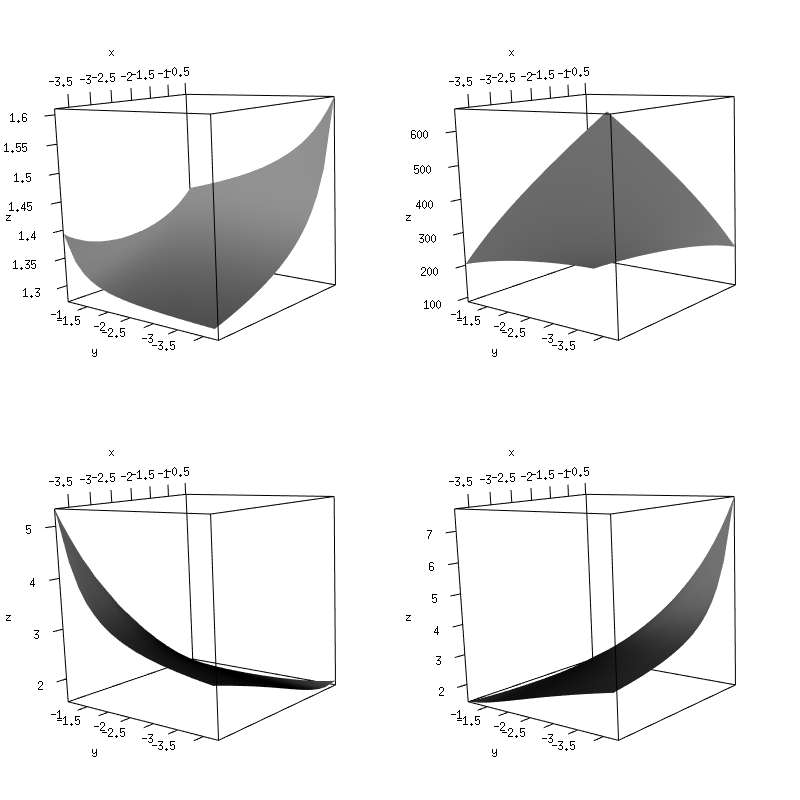}
\caption{Performance of the optimal test for $\theta_0=0.2$ vs. $\theta_1=0.3$:\\
$R(\psi^*)$ and $N(\theta_*;\psi^*)$ (above), $R_0(\psi^*)$ and $R_1(\psi^*)$ (below)}
\end{figure}

\begin{figure}
\includegraphics[scale=0.5]{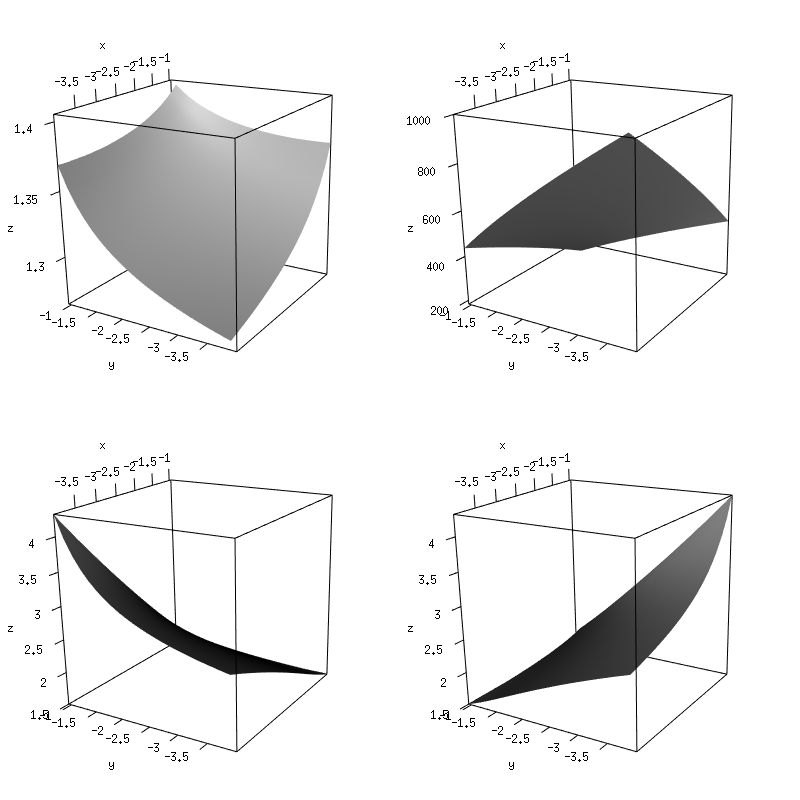}
\caption{Performance of the optimal test for $\theta_0=0.45$ vs. $\theta_1=0.55$:\\
$R(\psi^*)$ and $N(\theta_*;\psi^*)$ (above), $R_0(\psi^*)$ and $R_1(\psi^*)$ (below) }

\end{figure}

\section{Discussion and conclusions}

\subsection{Analysis of the numerical results}

 Tables 1 -- 5 are convenient for analyzing the relative  efficiency $R(\psi^*)$ of the optimal tests with respect to the fixed-sample-size test. It is clearly seen that it does not vary much within the whole range of $\alpha$ and $\beta$ computed. Its lowest value is about 1.3 and is attained at the minimum values of $\alpha $ and $\beta$ considered. There is a clear tendency of the relative efficiency decreasing with $\alpha=\beta\to 0$. It would be interesting to have a meaningful theoretical lower bound for its limiting value, but we doubt it could ever be obtained. The only fact  we know is that the relative efficiency by definition can not drop below 1.

The relative efficiency $QR(\psi^*)$ based on the 0.99-quantile of the optimal Kiefer-Weiss test behaves quite well maintaining  the approximate level of 0.6 to 0.8, for practical values of $\alpha=\beta$.

The relative efficiency of the SPRT based on the average sample number evaluated at $\theta_*$ drops to approx. 0.7 -- 0.8 for lower levels of $\alpha$ and $\beta$, which could still be considered tolerable, at least it is comparable to the efficiency 1.3 of the optimal Kiefer-Weiss test. What is catastrophic for the SPRT is its low efficiency $QR(W)$ based on the 0.99-quantile of the sample number distribution, under $\theta_*$, which is as low as approx. 0.3 -- 0.2, meaning the 0.99-quantile can reach a 3 -- 5 times higher level than the fixed sample size.

The relative efficiency based on the average sample number calculated at $\theta_*$ is important as a guideline for constructing optimal tests, but has only a limited importance for practical applications:  the operating characteristic function evaluated  at $\theta^*$  is always about 0.5, thus, under this circumstance, the optimal test  is as useful as  flipping a coin.
The "minimax" Kiefer-Weiss criterion is more like some "robustness" principle for constructing tests.
A more in-depth  discussion of this aspect of the Kiefer-Weiss problem and related questions   can be found in \cite{fauss2020fading}.

What really matters for the performance of a test is its behaviour under $H_0$ and $H_1$. To analyze this, we define two other characteristics: $R_0(\psi^*)=FSS/N(\theta_0;\psi^*)$ and  $R_1(\psi^*)=FSS/N(\theta_1;\psi^*)$. The behaviour of the three characteristics, $R$, $R_0$ and $R_1$, can be observed on Figures 1 -- 3 as a function of $\alpha$ and $\beta$ (on the logarithmic scale).

It is clearly seen that $R(\psi^*)$ keeps maintaining its level at approx. 1.3 -- 1.5, now in the whole range of $\alpha$ and $\beta$, not only for equal ones.

The relative efficiency under $H_0$, $R_0$, tends to have higher values for very asymmetric cases of small $\alpha$ and large $\beta$ and lower values for large $\alpha$ and small $\beta$, being within  1.8 -- 2.4 in the vicinity of the diagonal $\alpha=\beta$.
 $R_1$ behaves nearly  symmetrically to $R_0$ with respect to the diagonal $\alpha=\beta$. In the symmetric case $\theta_0=1-\theta_1$, $R_0$ and $R_1$ are perfectly symmetric to each other.

In general, the optimal tests in the Kiefer-Weiss setting maintain good levels of efficiency, requiring, under $H_0$ and $H_1$,
on the average 1.8 -- 2.4 times fewer observations than the best fixed-sample-size test (for $\alpha\approx\beta$),  and 1.3 -- 1.5 times fewer, in the worst case when the hypothesis is misspecified.

\subsection{Further work}

There is a lot of work can be done on the basis of the results we obtained here.

The most immediate is to study the efficiency of Lorden's 2-SPRT with respect to the optimal Kiefer-Weiss test in the case of Bernoulli observations. The symmetric normal case is analyzed numerically in \cite{Lorden76}.

 For the Bernoulli case, we now have the software for exact calculation of the optimal Kiefer-Weiss tests and all their characteristics. In addition, 
the subroutines for  calculation of the operating characteristic and the average sample number we have in our package,  can be used for 2-SPRT as well (in fact, they are applicable to any sequential test). As an alternative, one can make use of the  
algorithm described by \cite{Causey} specifically for the 2-SPRT. Thus, we are in a position to evaluate the efficiency of the 2SPRT with respect to the optimal test in the Kiefer-Weiss problem, for the Bernoulli model. The exact results may shed light on the precision of the general asymptotic results of \cite{Lorden76}, \cite{Huffman1983} in this particular case.

Another direction for future work is the implementation of the proposed method for other distribution families. Without any doubt, this can be done for discrete one-parametric Koopman-Darmois families, directly applying Proposition 1 for obtaining solutions to the modified Kiefer-Weiss problem. Applications to continuous distributions seem to be also viable using quadrature formulas for integrals in \eqref{14a_2}, similar to those suggested by \cite{Li2016} (see Section II.E) for computation of  the SPRT characteristics.

At last, a relationship with a continuous-time model can be anticipated, when the hypotheses get close ($\theta_1\to\theta_0$ in our case). In view of invariance principles, the limiting characteristics of the optimal tests in the Bernoulli case are expected to be those corresponding to the Kiefer-Weiss problem for Wiener process with a linear drift \citep[cf.][]{Lai1973}. Therefore, our results will provide an approximation to the limiting continuous-time problem. Because the theoretical solution to the latter is known  for $\alpha=\beta$ 
\citep[see][]{Zhitlukhin}, numerical estimation of the rate of convergence can be obtained in this case. For $\alpha\not=\beta$, our approximations may shed light on the solution of the Kiefer-Weiss problem for the Wiener process.

\section*{Acknowledgements}

A. Novikov (Mexico) is grateful to SNI by CONACyT,  Mexico, for a partial support for his work.\\
The work of A. Novikov (Russia) was partially supported by Russian Foundation for Basic Researh grant 19-29-01058.\\
F. Farkhshatov thanks CONACyT, Mexico, for  scholarship for his doctoral studies.\\

\end{document}